\documentclass[24pt]{article}
\usepackage{amssymb}
\usepackage{epsfig,caption,graphicx,eepic,epic}
\usepackage{pstricks}
\usepackage{amssymb,amsmath}
\usepackage{multido}
\usepackage{array}
\begin{document}
\def\be{\begin{equation}}
\def\ee{\end{equation}}
\def\ba{\begin{eqnarray}} 
\def\ea{\end{eqnarray}}
\def\nn{\nonumber}

\newcommand{\bbf}{\mathbf}   
\newcommand{\rrm}{\mathrm}
\title{Divisibility and coherence properties of an open quantum system: role of the interaction with the environment}

\author{Tarek Khalil$^{a}$
\footnote{E-mail address: tkhalil@ul.edu.lb}\\ 
and\\
Jean Richert$^{b}$
\footnote{E-mail address: j.mc.richert@gmail.com}\\ 
$^{a}$ Department of Physics, Faculty of Sciences(V),\\
Lebanese University, Nabatieh,
Lebanon\\
$^{b}$ Institut de Physique, Universit\'e de Strasbourg,\\
3, rue de l'Universit\'e, 67084 Strasbourg Cedex,\\ 
France} 
\date{\today}   
\maketitle 
\begin{abstract}
In the present work we recall and extend the results of previous work concerning the time evolution of  open quantum systems. We show how general properties of such systems are related to their structure properties, those of their environments and the interaction between the systems and the environments.  
\end{abstract}
\maketitle
PACS numbers: 02.50.Ey, 02.50.Ga, 03.65.Aa, 03.65.Yz
\vskip .2cm
Keywords: open quantum systems, divisibility and coherence criteria.\\

\section{General introduction}

What are the criteria which allow to classify open quantum systems with respect to their behaviour in the presence of an environment? Clearly, the evolution of two interacting systems can only depend on their structure and the structure of the interaction which governs their coupling. In the present investigations we aim to find out up to what extent the dynamical properties, i.e. the dynamical evolution  of an open quantum system can be understood by the commutation properties of the Hamiltonians which describe the system, its environment and their interaction.

The analysis contains the following items. First we address the question of the Markovian and time divisibility (semi-group) property of such systems. In a second part we analyze the conditions under which a finite open system behaves coherently or incoherently during its time evolution.\\

The structure of the paper is the following. In the first part we recall in section 2 some mathematical definitions of a stochastic Markov process and show how the concept leads to the description of the evolution of an open quantum system. There it appears that the Markovian property stems from the existence of different time scales which are related through the nature of the interaction between the system and the environment. In section 3 we concentrate on  the relation between a Markov process and the so called divisibility property, an essential concept in the description of open systems since it fixes the conditions under which the evolution is governed or not by a unique time variable. Section 4 we summarize the results. 

In the second part we address the concept of coherence and decoherence again in relation to the properties of the Hamiltonians which describe the dynamics of the system coupled to its environment.    
Section 5 is devoted to the determination of the conditions under which the evolution of an open system is either coherent or decoherent, showing how the quantum aspect of the phenomenon or its absence can be related to the properties of the different Hamiltonians which enter the description. Examples are developed and discussed in section 6 and conclusions are drawn in section 7. Section 8 gives a summary of  the results and presents general conclusions. Details of calculations are developed in Appendices A to E.

\section{Markov processes and divisibility}

The interaction between open quantum systems and their environment generates a response of the system which is due to the coupling between the two parts. An important amount of work on this subject concerning the evolution of such systems has been developed over several decades. Among the most recent ones one may consult refs.~\cite{riv1,pol}.

In mathematics a Markov process is a stochastic process in which a given event at a fixed time depends only on the preceding event, hence the existence of former events are forgotten. Formally the conditional probability for an event to be in a given state at a fixed time $t$ depends only on the state of the event at the closest time preceding $t$.

In physical systems like open quantum systems the action of the environment on the system generally induces an action which survives over a certain interval of time, the memory time which may be finite or not~\cite{ad,fle1}. The process is said to be non-Markovian if this time is finite. The process is said to be Markovian if the memory time tends to zero, i.e. if the correlations induced by the interaction between the system and its environment survive sequentially and independently over short intervals of time which tend to zero. Physically speaking the typical time scale over which the system evolves in time is much larger that the time scale which characterizes the evolution of the environment with which it interacts. Markov processes are described in terms of master equations which contain a unique time coordinate.

If the time  scale constraints defined above are no longer obeyed another regime sets in. There appears a delay in the response of the system to the action of the environment and the evolution of the system in a time interval $[t_{1}$, $t_{2}]$ depends on the history of the coupled system and its environment. The master equation which governs the evolution of the system depends then on more than one time variable, memory effects set in. The evolution of the system gets non-Markovian. 

A rigorous treatment of the two-time memory kernel in a master equation description of the evolution of a system in interaction with an external system is a delicate task which involves the problem of time hierarchies~\cite{sp}. The explicit structure of such a kernel is different for each considered system. It is often tempting to approximate the expression of the master equation which governs the evolution of the system by a phenomenological kernel of Markovian nature~\cite{gk,gl,lzb,sr}. Tests which measure the validityof the approximations have been proposed. They rely on positivity properties of the master equation which governs the evolution of the open system and  allow to distinguish between Markovian and non-Markovian behaviour ~\cite{hal}.

Recently two aspects of the memory problem have been re-examined in different contributions. The first one concerns the effective derivation of non-Markovian transport equations~\cite{fle1,hb,vb,ar,fb,sl,ss,mw,bv}, the second one the characterization of the deviation of a process from a Markovian behaviour~\cite{blp,afp,pcz,hpb,ck} by means of a measure of the strength of memory effects.

The present contribution aims to show how the properties which characterize Markovian and non-Markovian systems can be derived from the commutation relations between the different Hamiltonian operators which describe the open system, its environment and the coupling between them. As a central result we show  the origin of the divisibility property of a Markovian system. This property may however also characterize systems which are not necessarily Markovian as defined above. Non-Markovian systems do not obey the divisibility property. We consider under what conditions it is possible to describe the evolution of these systems in terms of a master equation governed by a unique time variable.

\subsection{Mathematical definitions of a Markovian process}

We recall here some mathematical aspects of Markov processes and their relationship with the so called physical Markovian property in order to show the link with its relationship with its use for the description of physical systems. The original paper concerning this concept by A. A. Markov was published in russian in the Bulletin of the Mathematical and Physical Society of the University of Kazan, 1906. An english translation can be found in ~\cite{mark}. The mathematical concepts developed in the present section have been taken from ref.~\cite{oli}.\\  

{\bf 1) Random variables}

Consider a sample space $\Omega$ of possible outcomes of a random process ${w_{i}}$. Each outcome $i$ is an event. The assignation of a real number to each $w$ leads to a random variable $X(w)$, a single-valued real function of $w$ and $\Omega$ is the domain of $X$.\\

{\bf 2) Probabilities}

Consider a random variable $X$ and $x$ a fixed real number, $A_{X}$ the subset of $\Omega$ which consists of all real sample points to which $X$ assigns the number $x$ 

\ba
A_{X}=[w | X(w)=x]=[X=x]
\label{eq1}
\ea

Because $A_{x}$ is an event it will have a probability $p=P(A_{x})$. One defines a cumulative distribution function as
  
\ba
F_{X}(x)=P[X \leq x] 
\label{eq2}
\ea
 with $x$ in the interval $[- \infty, + \infty]$.\\  
  
{\bf 3) Stochastic or random process}

Consider a set of random variables depending on a continuous variable $t$. Define $X(t,w)$ as a collection of time functions for a fixed value of $w$.

A stochastic or random process is a family of random variables $[X(t,w)]$ defined over a given parameter set $T$ indexed by $t$. In the following the fixed event parameter $w$ will be left out in the notations.\\

{\bf 4) Markov process and strong Markov process}

\begin{itemize}

\item A stochastic process $[X(t)]$ where $t$ belongs to a continuous ensemble $T$ is called a 1st order Markov process if for a sequence $[t_{0}, t_{1},...,t_{n}]$ the conditional cumulative distribution function $F_{X}$ of
$X(t_{n})$ for a given sequence $X(t_{0})$, $X(t_{1})$,..., $X(t_{n-1})$ depends only on $[X(t_{n-1})]$:

\ba
P[X(t_{n}) \leq x_{n} | X(t_{n-1})=x_{n-1}, X(t_{n-2})=x_{n-2},...,X(t_{0})=x_{0}]=                                                        
\notag\\
P[X(t_{n}) \leq x_{n} | X(t_{n-1})=x_{n-1}]
\label{eq3}
\ea  
  
\item The process is strong if $[X(t+s)-X(t), s \geq 0]$  has the same distribution as the process
$[X(s),s \geq 0]$ and is independent of the process $[X(s),0 \leq s \leq t]$ i.e., if the process is known at time $t$ the probability law of the future change of state of the process will be determined as if the process started at time $t$, independently of the history of the process between $t=0$ and $t$.\\

\end{itemize}

One can interpret the Markovian property of a system $S$ which evolves stochastically as a loss of memory of the system over an arbitrarily short interval of time. Hence the evolution is governed by a process such that this evolution at any arbitrary time $t$ depends only on time intervals $t-dt,t$, independently of its evolution in the past. The strong limit of the process shows that the evolution is invariant under time translation, it depends only on the time interval between two random events and not of the initial time at which the process is observed.

The consequence of the specific time dependence of the Markov assumption applied to physical systems is the divisibility property which will be shown in the following section. We follow the development proposed in ref.~\cite{har}.  

\subsection{Physical application: conditions for the existence of a Markovian master equation}.

Consider an open quantum system $S$ coupled to its environment $E$. In general $S$ is described by its time-dependent density operator $\hat \rho_{S}(t)$ and can be obtained as the solution of a master equation. The evolution can be followed in terms of the differential equation

\ba
\frac{d\hat \rho_{S}(t)}{dt}=lim_{\tau \rightarrow \to 0} \frac{\hat \rho_{S}(t+\tau)-\hat \rho_{S}(t)}{\tau}
\label{eq4}
\ea
where $\hat \rho_{S}(t+\tau)=\hat L_{t,\tau} \hat \rho_{S}(t)$, 
$\hat L_{t,\tau}$ being the evolution operator from $t$ to $t+\tau$. 

Since $S$ is coupled to $E$ the total system $S+E$ is described by a density operator $\hat \rho_{SE}(t)$ whose general expression can be written   

\ba
\hat \rho_{SE}(t)=\hat \rho_{S}(t)\hat \rho_{E}(t)+\delta\hat \rho_{SE}(t)
\label{eq5}
\ea
and $\hat \rho_{S}(t)=Tr_{E}[\hat \rho_{SE}(t)]$. 
The operator $\delta\hat \rho_{SE}(t)$ is generated by the interaction between $S$ and $E$. It perturbs the free evolution of $S$ and may induce retardation effects in the process, hence need the introduction of a time $s \leq t$ which describes this effect in the evolution of the system. Eq.(4) will  lead to a Markovian master equation if two conditions are satisfied:

\begin{itemize}

\item (a) $\delta\hat \rho_{SE}(t)/\hat \rho_{SE}(t) \ll 1$ 

\item (b) $\| \hat \rho_{E}(t) \| \simeq \bar \rho_{E} $ 

\end{itemize}
where $\bar \rho_{E}$ is a constant density. Relation (a) originates from the fact that the correlations induced by the coupling have to be negligible, (b) expresses the fact that $E$ is stationary,  not influenced by $S$. As a consequence $L_{t,\tau}$ must be independent of $t$ so that $E$ does not depend on $S$ in earlier times. We shall come back to this point in the sequel.

\subsection{Markov process}

The properties mentioned above can be realized under specific conditions. If the environment is a large system insensitive to the presence of $S$ and the extension of its spectrum $\Delta_{E}$ is large the decay time $\tau_{E}=\hbar/\Delta_{E}$ (the time over which the correlations generated by the coupling interaction $\hat H_{SE}$ between $S$ and $E$ survive) is small. The time $\tau_{E}$ is the characteristic memory time of $E$ over which the correlations in $E$ survive. During this time the phases of $S+E$ change by an amount of the order of $\tau_{E}||\hat H_{SE}||/\hbar$. These phases have to be small in order to verify the conditions given above. Indeed, it comes that   

\ba
\delta\hat \rho_{SE}(t)=O(||\hat H_{SE}||^{2}\tau_{E}^{2}/\hbar^{2})
\notag\\
\hat \rho_{E}(t)=\bar \rho_{E}+O(||\hat H_{SE}||^{2}\tau_{E}^{2}/\hbar^{2})
\label{eq6}
\ea
where $||\hat H_{SE}|$ is the strength of the interaction. The inequality $||\hat H_{SE}||\tau_{E}/\hbar \ll 1$ qualifies the Markovian property of the process: the time interval over which the system keeps the memory of its coupling to the environment and the coupling between $S$ and $E$ have to be small.   

\subsection{Time scales}

The system $S$ is characterized by a typical evolution time $\tau_{S}$. Consider the evolution of the phases of $S+E$ as a succession of phases which accumulate coherently over time intervals $\tau_{E}$. The process operates as a random walk over large time intervals $t$ during which the phases $\Phi(t)$ add up quadratically as

\ba
\Delta^{2} \Phi(t \gg \tau_{E}) \sim (||\hat H_{SE}||\tau_{E}/\hbar)^{2} t/\tau_{E}=t/\tau_{S}
\label{eq7}
\ea
the time $\tau_{S}=\hbar^{2}/(||\hat H_{SE}||^{2}\tau_{E})$ is the typical time over which the system $S$ evolves. From this expression one can see that the time $\tau_{S}$ is much longer than $\tau_{E}$ when the process is Markovian. One can rewrite expressions of eq.(6)

\ba
\delta\hat \rho_{SE}(t)=O(\tau_{E}/\tau_{S})
\notag\\
\hat \rho_{E}(t)=\bar \rho_{E}+O(\tau_{E}/\tau_{S})
\label{eq8}
\ea

\section{Divisibility and Markov processes}

We derive here the origin of the divisibility property of a Markov process. We look for the condition under which the time evolution of the density operator $\hat \rho_{S}(t)=Tr_{E}[\hat \rho_{SE}(t)]$ is given by
 
\ba
\hat\rho_{S}(t)=\hat \Phi(t,t_{0}) \hat\rho_{S}(t_{0})
\label{eq9}
\ea
where $\hat \Phi(t,t_{0})$ is the evolution operator in $S$ space.

The master equation derived above describes a Markovian system because its derivation relies on the memoryless behaviour of the correlations which characterize the coupling of the system to its stationary environment. Indeed, in the limit where $\tau_{E}$ gets very small compared to the characteristic evolution time of $S$ the time correlations 

\ba 
C(t,t')=\langle \hat H_{SE}(t) \hat H_{SE}(t'=t+\tau_{E}) \rangle  \propto \delta(t-t')
\label{eq10}
\ea
hence at a finite time $t' \neq t$ the system has lost the memory of its coupling to the environment at time $t$, the behaviour of the system at $t'$ is independent of its behaviour at $t$.  
 
As a consequence the master equation  which governs the evolution of $\hat \rho_{S}(t)$ at each time $t$ depends on a single time variable. The density operator $\hat\rho_{S}(t_{2})$ will be related to 
$\hat\rho_{S}(t_{1})$ $(t_{2}>t_{1})$ by 
$\hat \rho_{S}(t_{2})= \hat \Phi(t_{1},t_{2})\hat \rho_{S}(t_{1})$. Then for any further time interval $[t_{2},t_{3}]$ one will get                                    

\ba 
\hat \rho_{S}(t_{3})= \hat \Phi(t_{3},t_{2})\hat \Phi(t_{2},t_{1}) \hat \rho_{S}(t_{1})
\label{eq11}
\ea 
where $\hat \Phi(t',t)$ is the evolution operator of the open system ~\cite{gl,hb,vb,ar}.\\

Relation (11) defines the divisibility property. It comes as a property of Markov systems which are governed by time scale considerations. The question which comes next is to know whether this property is restricted to this type of systems, hence if divisibility is equivalent to Markovianity: are there other physical conditions which lead to this property? The question is answered below. 

\subsection{Divisibility in open quantum systems}

We consider now the case where an open system does not necessarily fulfill the time scale conditions which lead to a Markovian system but nevertheless obeys the divisibility condition. \\

1) General expression of the total density operator

Consider a system $S$ characterized by a density operator  $\hat\rho_{S}(t)$ which evolves in time from $t_{0}$ to $t$
under the action of the evolution operator $\hat \Phi(t,t_{0})$

\ba
\hat\rho_{S}(t)=\hat \Phi(t,t_{0})\hat\rho_{S}(t_{0})
\label{eq12}
\ea

 At the initial time $t_{0}$  the system $S$ is supposed to be decoupled from its environment and characterized by the density operator

\ba
\hat\rho_{S}(t_{0})= \sum_{i_{1},i_{2}}c_{i_{1}}c_{i_{2}}^{*}|i_{1}\rangle \langle i_{2}|
\label{eq13}
\ea
and the environment $E$ by

\ba
\hat\rho_{E}(t_{0})= \sum_{\alpha_{1},\alpha_{2}}d_{\alpha_{1},\alpha_{2}}|\alpha_{1}\rangle \langle \alpha_{2}|
\label{eq14}
\ea
where $|i_{1}\rangle, |i_{2}\rangle $ and $|\alpha_{1}\rangle, |\alpha_{2}\rangle$ are orthogonal states in  
$S$ and $E$ spaces respectively, $c_{i_{1}},c_{i_{2}}$ normalized amplitudes and $d_{\alpha_{1},\alpha_{2}}$ weights such that $\hat\rho_{E}^{2}(t_{0})=\hat\rho_{E}(t_{0})$.

At time $t>t_{0}$ the reduced density operator in $S$ space is $\hat\rho_{S}(t)=Tr_{E}[\hat\rho(t)]$ where $\hat\rho(t)$ is the density operator of the total system $S+E$. It can be written as~\cite{vlb}

\ba
\hat\rho_{S}(t)=\sum_{i_{1},i_{2}}c_{i_{1}}c_{i_{2}}^{*}\hat\Phi_{i_{1},i_{2}}(t,t_{0})
\label{eq15}
\ea
with

\ba
\hat\Phi_{i_{1},i_{2}}(t,t_{0})=\sum_{j_{1},j_{2}}C_{(i_{1},i_{2}),(j_{1},j_{2})}(t,t_{0})|j_{1}\rangle_{S}\langle j_{2}|
\label{eq16}
\ea

where the super matrix $C$ reads

\ba
C_{(i_{1},i_{2}),(j_{1},j_{2})}(t,t_{0})=\sum_{\alpha_{1},\alpha_{2},\gamma}d_{\alpha_{1},\alpha_{2}}
U_{(i_{1}j_{1}),(\alpha_{1}\gamma)}(t,t_{0}) U_{(i_{2}j_{2}),(\alpha_{2}\gamma)}^{*}(t,t_{0})
\label{eq17}
\ea

and

\ba
U_{(i_{1}j_{1}),(\alpha_{1}\gamma)}(t,t_{0})=\langle j_{1}\gamma|\hat U(t,t_{0})|i_{1} \alpha_{1}\rangle
\notag\\
U^{*}_{(i_{2}j_{2}),(\alpha_{2}\gamma)}(t,t_{0})=\langle i_{2} \alpha_{2}|\hat U^{*}(t,t_{0})|j_{2} \gamma \rangle
\label{eq18}
\ea

The evolution operator reads $\hat U(t,t_{0})=e^{-i\hat H(t-t_{0})}$ where $\hat H$ is the total Hamiltonian in 
$S+E$ space and the super matrix $C$ obeys the condition 
$\lim_{t\rightarrow t_{0}}C_{(i_{1},i_{2}),(j_{1},j_{2})}(t,t_{0})=\delta_{i_{1},i_{2}} \delta_{j_{1},j_{2}}$.\\

In the present formulation the system is described in terms of pure states. The results which will be derived below remain valid if the initial density operator at the initial time is composed of mixed states 
$\hat\rho_{S}(t_{0})=\sum_{i_{1},i_{2}}c_{i_{1}i_{2}}|i_{1}\rangle _{S}\langle i_{2}|$.\\

2) Divisibility constraint: particular case

Consider a system which obeys the divisibility criterion ~\cite{ar,has,bv} 

\ba
\hat\rho_{S}(t,t_{0})=\hat \Phi(t,\tau)\hat \Phi(\tau,t_{0})\hat\rho_{S}(t_{0})
\label{eq19}
\ea
For $\tau$ in the interval $[t_{0},t]$. Eq.(18) expresses the fact that $\hat T$ which acts between $t_{0}$ 
and $t$ obeys the  divisibility criterion.

The problem is now to find conditions under which the general expression of $\hat\rho_{S}(t,t_{0})$ obeys the divisibility constraint fixed by Eq.(20) at any time $t>t_{0}$. For this to be realized the following relation must be verified by the super matrix $C$

\ba
C_{(i_{1},i_{2}),(k_{1},k_{2})}(t,t_{0})= \sum_{j_{1},j_{2}}C_{(i_{1},i_{2}),(j_{1},j_{2})}(t_{s},t_{0})
C_{(j_{1},j_{2}),(k_{1},k_{2})}(t,t_{s})
\label{eq20}
\ea
The explicit form of this equation is worked out in Appendix A. Writing out explicitly the r.h.s. and l.h.s. of Eq.(20) in terms of the expression of $C$ given in Eq.(17) for fixed values of $i_{1}$ and $i_{2}$ one finds from inspection that a sufficient condition for this to be realized is obtained if there is a unique state $|\eta\rangle$ in $E$ space with $d_{\eta,\eta}=1$. This is in agreement with ref.~\cite{sti}.\\  

3) Divisibility and properties of the coupling between $S$ and $E$ space

The former result can be extended to a more general case. To see this we introduce the explicit expression of the master equation which governs an open quantum system in a time local regime which can be put in the general form ~\cite{gk,gl}

\ba
\frac{d}{dt}\hat\rho_{S}(t)=\sum_{n}\hat L_{n}(t)\hat\rho_{S}(t)\hat R_{n}^{+}(t)
\label{eq21}
\ea
where $\hat L_{n}(t)$ and $\hat R_{n}(t)$ are time local operators.\\

The matrix elements of the density operator $\hat\rho_{S}(t)$ given by Eqs. (15-18) read

\ba
\hat\rho_{S}^{j_{1}j_{2}}(t)=\sum_{i_{1}i_{2}}c_{i_{1}}c^{*}_{i_{2}}\sum_{\alpha_{1},\alpha_{2}\gamma}d_{\alpha_{1},\alpha_{2}}\langle j_{1}\gamma|\hat U(t,t_{0})|i_{1} \alpha_{1}\rangle
\notag\\
\langle i_{2} \alpha_{2}|\hat U^{*}(t,t_{0})|j_{2} \gamma \rangle
\label{eq22}
\ea
Taking its time derivative leads to two contributions to the matrix elements of the operator 

\ba
\frac{d}{dt}\rho_{S1}^{j_{1}j_{2}}(t)=(-i)\sum_{i_{1}i_{2}}c_{i_{1}}c^{*}_{i_{2}}\sum_{\alpha_{1},\alpha_{2}}d_{\alpha_{1},\alpha_{2}}
\sum_{\beta \gamma k_{1}}\langle j_{1}\gamma|\hat H|k_{1} \beta \rangle
\notag\\
\langle k_{1}\beta|e^{-i\hat H(t-t_{0})}|i_{1} \alpha_{1} \rangle \langle i_{2}\alpha_{2}|e^{i\hat H(t-t_{0})}|j_{2} \gamma \rangle 
\notag\\
\frac{d}{dt}\rho_{S2}^{j_{1}j_{2}}(t)=(+i)\sum_{i_{1}i_{2}}c_{i_{1}}c^{*}_{i_{2}}\sum_{\alpha_{1},\alpha_{2}}d_{\alpha_{1},\alpha_{2}}
\sum_{\beta \gamma k_{2}}\langle j_{1}\gamma|e^{-i\hat H(t-t_{0})}|i_{1} \alpha_{1} \rangle  
\notag\\
\langle i_{2}\alpha_{2}|e^{i\hat H(t-t_{0})}|k_{2} \beta \rangle \langle k_{2}\beta|\hat H|j_{2} \gamma \rangle 
\label{eq23}
\ea
and                  

\ba
\frac{d}{dt}\hat\rho_{S}^{j_{1}j_{2}}(t)=\frac{d}{dt}\rho_{S1}^{j_{1}j_{2}}(t) + \frac{d}{dt}\rho_{S2}^{j_{1}j_{2}}(t)
\label{eq24}
\ea
From the explicit expression of the density operator matrix element given by Eqs. (23-24) it can be seen that the structure of the master equation above can only be realized if $|\beta \rangle=|\gamma \rangle$. Three solutions can be found:

\begin{itemize}

\item There is a unique state $|\gamma \rangle$ in $E$ space. This result has already been seen on the expression of the density operator above.

\item The density operator $\hat\rho_{S}(0)$ is diagonal in $S$ space with equal amplitudes of the states 
and the states in $E$ space are equally weighed, $\hat \rho_{E}=\sum_{\alpha}d_{\alpha,\alpha}|\alpha\rangle\langle\alpha|$, $d_{\alpha,\alpha}=1/N$ where $N$ is the number of states in $E$ space. See details in Appendix B. These states called maximally coherent states have been introduced recently in a study of quantum coherence ~\cite{bau}.

\item The environment stays in a fixed state $|\gamma \rangle$, i.e. if the
Hamiltonian $\tilde H=\hat H_{E}+\hat H_{SE}$ is diagonal in a basis of states in which $\hat H_{E}$ is diagonal. Then the density operator is the sum $\hat\rho_{S}(t,t_{0})=\sum_{\gamma}\hat\rho_{S \gamma}(t,t_{0})$ see Appendix C. In this case the commutation relation  $[\hat H_{E},\hat H_{SE}]=0$ is the central result.

\end{itemize}

All three conditions are sufficient to insure the structure of the  r.h.s. of Eqs.(21-23), the last one being the most general one. We shall come back to the physical immplication of the last point in the sequel. 

\subsection{Memory effects and absence of divisibility: two-time approach}

Using the projection formalism ~\cite{nak,zwa,ck1} and the expression developed in section 3.1 the density operator of the total system $S+E$ reads 

\ba
\hat \rho(t,t_{0})=\sum_{i_{1},i_{2}}c_{i_{1}}c_{i_{2}}\sum_{\alpha}d_{\alpha \alpha}U(t,t_{0})
|i_{1} \alpha \rangle \langle i_{2} \alpha| U^{+}(t,t_{0})
\label{eq25}
\ea
Without loss of generality we write the expression of $\hat \rho(t,t_{0}$ in a basis of states in which $\hat H_{E}$ is diagonal.

Introduce projection operators $\hat P$ and $\hat Q$  such that

\ba
\hat P \hat \rho(t,t_{0})=\sum_{k=1}^{n}|\gamma_{k}\rangle \langle \gamma_{k}|\hat \rho(t,t_{0})
\notag\\
\hat Q \hat \rho(t,t_{0})=\sum_{l=n+1}^{N}|\gamma_{l}\rangle \langle \gamma_{l}|\hat \rho(t,t_{0})
\label{eq26}
\ea
where $N$ is the total finite or infinite number of states in $E$ space and $\hat P+\hat Q=\hat I$ 
where $\hat I$ is the identity operator.

The evolution of the density operator is given the Liouvillian equation
\ba
\frac { d \hat \rho(t,t_{0})}{ dt}=\hat L(t) \hat \rho(t,t_{0})=-i[\hat H,\hat \rho(t,t_{0})]
\label{eq27}
\ea

Projecting this equation respectively on $\hat P$ and $\hat Q$ subspaces leads to a set of two coupled  equation

\ba
\frac{d \hat P \hat \rho(t,t_{0})}{dt}= \hat P \hat L(t)\hat P \hat \rho(t,t_{0})+
\hat P \hat L(t)\hat Q \hat \rho(t,t_{0}) (a)
\notag\\
\frac{d \hat Q \hat \rho(t,t_{0})}{dt}= \hat Q \hat L(t)\hat Q \hat \rho(t,t_{0})+
\hat Q \hat L(t)\hat P \hat \rho(t,t_{0}) (b)
\label{eq28}
\ea

Solving formally the second equation gives
\ba
\hat Q(t)\hat \rho(t,t_{0})=e^{\hat Q(t)\hat L(t)t}\hat Q(t)\hat \rho(t=t_{0})+ \int ^{t}_{t_{0}}
dt'e^{\hat Q(t')\hat L(t')t'}\hat Q \hat L(t')\hat P \hat \rho(t',t_{0})
\label{eq29}
\ea
and inserting into the first equation one obtains
\ba
\frac{d \hat P \hat \rho(t,t_{0})}{dt}= \hat P \hat L(t)\hat P \hat \rho(t,t_{0})+\hat P \hat L(t)
\int ^{t}_{t_{0}}dt'e^{\hat Q(t')\hat L(t')(t-t')}\hat Q \hat L(t')\hat P \hat \rho(t',t_{0})
\label{eq30}
\ea  

This first order two-time integro-differential equation reduces to an ordinary one-time differential equation if $\hat P$ subspace reduces to a single state or, more generally, if $[\hat H_{E},\hat H_{SE}]=0$.
In that case Eq.(28a) reads 

\ba
\frac{d \hat P \hat \rho(t,t_{0})}{dt}=i\hat P[\hat P \hat \rho(t,t_{0}),\hat H]+i\hat P
[\hat Q \hat \rho(t,t_{0}),\hat H]
\label{eq31}
\ea  
Developing $\hat H=\hat H_{S}+\hat H_{E}+\hat H_{SE}$ and inserting on the r.h.s. leads to 

\ba
\hat P [\hat Q \hat \rho(t,t_{0}),\hat H]=0
\label{eq32}
\ea
because $\hat H_{SE}$ is such that $\hat P \hat H_{SE} \hat Q=0$. Hence there remains 

\ba
\frac{d \hat P \hat \rho(t,t_{0})}{dt}=i\hat P[\hat P \hat \rho(t,t_{0}),\hat H]
\label{eq33}
\ea
which is local in time and leads to the divisibility property. This result is again in agreement with the results obtained above and also with ref.~\cite{ck}.
It shows that the evolution of the density operator $\hat \rho_{S}(t,t_{0})=Tr_{E}\hat \rho(t,t_{0})$ in $S$ space is governed by the local time  $t$.\\


\subsection{Memory effects and absence of divisibility: one-time approach}

Divisibility is obtained if $[\hat H_{E},\hat H_{SE}]=0$ in an basis of states in which $\hat H_{E}$ is diagonal. The violation of divisibility is realized when $\hat H_{SE}$ possesses
non-diagonal elements. Then the evolution of the density matrix is described by a master equation whose matrix elements for a fixed state $|\gamma\rangle$ in $E$ space depends on a unique time variable and takes the form 

\ba
\frac{d \hat \rho_{S\gamma}(t)}{dt}=(-i)[\hat H_{d}^{\gamma},\hat \rho_{S\gamma}(t)]+(-i)\sum_{\beta \neq \gamma}[\hat H_{nd}^{\gamma \beta}\hat M_{\beta \gamma}(t)-\hat M_{\gamma \beta}(t)
\hat H_{nd}^{\beta \gamma}]
\label{eq34}
\ea
where $\hat H_{nd}^{\gamma \beta}$ is a non diagonal matrix element in $E$ and $S$ space and 
$\hat M_{\gamma \beta}(t)$ the non diagonal matrix element which determines the time evolution of this term.

In the present formulation the master equation depends on a unique time variable although it describes a non divisible process.

\section{Summary}
  
The present analysis leads to the following conclusions:
  
\begin{itemize}

\item Markovian systems are governed by specific conditions concerning the time scales which characterize the system and its environment. Divisibility is an intrinsic property of these sytems. 

\item Divisibility is not restricted to Markovian systems. It is realized under specific conditions concerning the commutator properties of $\hat H_{E}$ and $\hat H_{SE}$, $[\hat H_{E},\hat H_{SE}]=0$, whatever the strength of the coupling interaction $\hat H_{SE}$. The master equation which governs the evolution of the matrix elements of the density matrix is diagonal in the environment space and contains only commutation operators in its r.h.s. 

\item The commutation relation is a sufficient condition for the existence of divisibility. It is not a necessary condition. 

\item The Markovian property induces the divisibility property. Divisibility does not necessarily induce a Markovian evolution. 

\item The master equation which governs the evolution of a non-Markovian system can take two different forms. It can either appear as an integro-differential equation with a memory kernel, i.e. contain two time coordinates. The time evolution of the density matrix can also contain a unique time coordinate but then it contains a further term due to the presence of non diagonal matrix contributions, 
$[\hat H_{E},\hat H_{SE}] \neq 0$. This is not the case if the system is Markovian.

\end{itemize} 
  
\section{Coherent and decoherent systems}

The second aspect which is closely related to the measure problem in quantum mechanics concerns the evolution of the components of the density operator in a fixed basis of states. Given a specific basis of states of the system the evolution can either lead to so called decoherence if the non diagonal matrix elements of this operator decrease to zero, coherent if this is not the case. The question we ask here concerns conditions under which an open quantum system behaves coherently or incoherently during its time evolution: can one find conditions which characterize  a coherent or a decoherent behaviour of quantum systems interacting with an external system such as a measuring device?  

Before investigating this point we consider first the case of a stationary system in order to exhibit the physical meaning of coherence, a specific property of quantum mechanical systems.

\subsection{Empirical considerations about interference effects in open quantum systems}

In the absence of interactions between two systems $S+E$ the density operator of the combined sytem 
can be written

\ba
\hat \rho_{SE}=\sum_{ij} c_{i}c^{*}_{j} |i \rangle \langle j| \sum_{\alpha \beta}d_{\alpha \beta}|\alpha \rangle \langle \beta|
\label{35}
\ea
in the notations introduced above. 

Taking the trace over the environment leads trivially to 
\ba
\hat \rho_{S}=\sum_{ij} c_{i}c^{*}_{j} |i \rangle \langle j|
\label{36}
\ea

If $|\phi \rangle= \sum_{i}c_{i}|i\rangle$ is the wave  function of the system  the probability that the system can go over to a state $|\psi \rangle=\sum_{i}b_{i}|i\rangle$ without affecting the environment is given by  

\ba
prob(\phi \to \psi)=|\sum_{i}c^{*}_{i}b_{i}|^{2}= \sum_{i}|c^{*}_{i}b_{i}|^{2}+
\sum_{ij;i\not=j}c^{*}_{i}c_{j}b^{*}_{j}b_{i}
\label{37}
\ea
which shows the expected interference effects due to the quantum nature of the system.  

Consider the case where at some time the interaction between $S$ and $E$ induces an effect such that $|i\rangle |\alpha \rangle \rightarrow |i\rangle |\alpha_{i} \rangle$ i.e. each state 
$|\alpha \rangle$ gets correlated with a specific state in $S$ space. Then the basis of states of the total system is ${|i\rangle |\alpha_{i}} \rangle$. Since $S+E$ must obey unitarity 

\ba
\langle i \alpha_{i}|j \alpha_{j}\rangle= \langle i|j \rangle \langle \alpha_{i}| \alpha_{j}\rangle
=\delta_{ij}
\label{38}
\ea

As a consequence the reduced density operator $\hat \rho_{S}$ will read

\ba
\hat \rho_{S}=Tr_{E}[\sum_{ij} c_{i}c^{*}_{j} |i \rangle \langle j| \sum_{\alpha_{i}\alpha_{j}} 
d_{\alpha_{i}\alpha_{j}}|\alpha_{i} \rangle \langle \alpha_{j}]=\sum_{i}|c_{i}|^{2}
|i\rangle \langle i|
\label{39}
\ea
and the non-diagonal terms of $\hat \rho_{S}$ due to interference effects have disappeared.

\subsection{Coherent and incoherent time evolution of open quantum systems}

The considerations of the preceding section will now be used as a guide to look for a theoretically founded relation able to decide whether an open quantum system behaves coherently or decoherently during its time evolution. In practice we use again the formalism developed in section 3.1. and look for the conditions under which decoherence can take place in $S$ during the time evolution of $S+E$.

The matrix elements of the projection of the total density operator 
$\hat\rho_{S+E}(t,t_{0})$ on $S$ space reads

\ba
\hat\rho_{S}^{j_{1}j_{2}}(t)=\sum_{i_{1}i_{2}}c_{i_{1}}c^{*}_{i_{2}}\sum_{\alpha_{1},\alpha_{2}\gamma}d_{\alpha_{1},\alpha_{2}}\langle j_{1}\gamma|\hat U(t,t_{0})|i_{1} \alpha_{1}\rangle
\notag\\
\langle i_{2} \alpha_{2}|\hat U^{*}(t,t_{0})|j_{2} \gamma \rangle
\label{eq40}
\ea
where $\hat U(t,t_{0})=exp[-i\hat H(t-t_{0})$ and $\hat H(t-t_{0})= \hat H_{S}+\hat H_{E}+\hat H_{SE}$. 

The behaviour of the open system with respect to coherence is governed by the structure of its coupling to the environment, in practice it is fixed by the commutator $[\hat H_{S},\hat H_{SE}]$. Indeed, if 
$\hat H_{S}$ commutes with $\hat H_{SE}$ one may expect the system which is prepared in a given state $j$ stays in this state during its time evolution. Refering to the empirical considerations developed above one may expect that in this case decoherence i.e. the decrease to zero of the non-diagonal contributions of the reduced density operator matrix should happen. We examine this case in the sequel.

If $[\hat H_{S},\hat H_{SE}]=0$ the reduced density operator can be written as 

\ba
\hat\rho_{S}^{j_{1}j_{2}}(t)=\sum_{i_{1}i_{2}}c_{i_{1}}c^{*}_{i_{2}}\sum_{\alpha_{1},\alpha_{2}\gamma}d_{\alpha_{1},\alpha_{2}}
\langle j_{1}|e^{-i\hat H_{S}(t-t_{0})}|j_{1}\rangle
\langle j_{1}\gamma|e^{-i(\hat H_{E}+\hat H_{SE})(t-t_{0})}|j_{1} \alpha_{1}\rangle
\notag\\
\langle j_{2}|e^{+i\hat H_{S}(t-t_{0})}|j_{2}\rangle
\langle j_{2}\alpha_{2}|e^{+i(\hat H_{E}+\hat H_{SE})(t-t_{0})}|j_{2} \gamma\rangle
\label{eq41}
\ea
 
Working out the matrix elements and noting that $\hat H_{S}|j \rangle = |\epsilon_{j}\rangle$ in a basis of states in which $\hat H_{S}$ is diagonal the diagonal matrix elements $j_{k},j_{k}$ of 
$\hat\rho_{S}(t)$ read
  
\ba
\hat\rho_{S}^{j_{k}j_{k}}(t)=\sum_{\alpha_{1},\alpha_{2}\gamma}d_{\alpha_{1},\alpha_{2}}
|c_{j_{k}}|^{2} 
\langle j_{k}\gamma|e^{-i(\hat H_{E}+\hat H_{SE})(t-t_{0})}|j_{k} \alpha_{1}\rangle
\notag\\
\langle j_{k}\alpha_{2}|e^{+i(\hat H_{E}+\hat H_{SE})(t-t_{0})}|j_{k} \gamma\rangle
\label{eq42}
\ea
One sees that the expression is completely symmetric in all indices and hence oscillates with time without dephasing.

The non-diagonal elements are given by
 
\ba
\hat\rho_{S}^{j_{k}j_{l}}(t)=\sum_{\alpha_{1},\alpha_{2}\gamma}d_{\alpha_{1},\alpha_{2}}
c_{j_{k}}c^{*}_{j_{l}}e^{-i(\epsilon_{j_{k}}-\epsilon_{j_{l})}(t-t_{0})}
\langle j_{k}\gamma|e^{-i(\hat H_{E}+\hat H_{SE})(t-t_{0})}|j_{k} \alpha_{1}\rangle
\notag\\
\langle j_{l}\alpha_{2}|e^{+i(\hat H_{E}+\hat H_{SE})(t-t_{0})}|j_{l} \gamma\rangle
\label{eq43}
\ea 
The matrix elements now induce different phases due to the fact that $j_{k} \not= j_{l}$, hence dephasing leading to decoherence may set in. At this stage it is not possible to conclude. Whether the non-diagonal elements decrease to zero and stay there depends on the explicit expression of the Hamiltonian 
$\hat H_{S}$ and $\hat H_{SE}$. There exist however specific cases for which an answer can be given. 
Below we quote two examples which exemplify different cases depending on the expression of $\hat H_{SE}$.

\section{Decoherent and coherent evolution of open systems: models}
 
\subsection{ Decoherent evolution}

Consider the model for which $\hat H=\hat H_{S}+\hat H_{E}+\hat H_{SE}$ with

 \begin{center}
\ba
\hat H_{S}=\omega_{0} \hat J_{z} 
\notag \\
\hat H_{E}=\sum_{k}\omega_{k} a^{+}_{k}a_{k}
\notag \\
\hat H_{SE}=\sum_{k}J_{k}(g_{k}a_{k}+g^{*}_{k}a^{+}_{k})
\label{eq44}
\ea 
\end{center} 
 
Here $\hat C_{H_{SE}}=[\hat H_{S},\hat H_{SE}]=0$ trivially. The model can be solved analytically ~\cite{rei,add}. Considering a two component spin system coupled to a bosonic reservoir it comes out that the nondiagonal components of the density operator $\hat \rho_{S}(t)$ read ~\cite{add} 
 
\ba
\rho_{i_{1}i_{2}}(t)= \rho^{*}_{i_{2}i_{1}}(t)=\rho_{i_{1}i_{2}}(0)e^{-\Gamma(t)}
\label{eq45} 
\ea 
where $\Gamma(t)$ is areal positive and decreasing function with time. 

\subsection{Coherent evolution}
 
a) Structure of the interaction and consequences\\

Consider now the case where the Hamiltonian $\hat H=\hat H_{S}+\hat H_{E}+\hat H_{SE}$ is constrained to $\hat C_{H_{SE}}=[\hat H_{S},\hat H_{SE}]=0$ like above but with a different form of the interaction which reads ~\cite{lid1,agr}
 
\ba
\hat H_{SE}=\sum_{\lambda} \hat g_{\lambda} \otimes \hat e_{\lambda}
\label{eq46}
\ea
where $\hat g_{\lambda}$ and $\hat e_{\lambda}$ are operators acting respectively in $\cal H_{S}$ and $\cal H_{E}$ Hilbert spaces such that the eigenvectors ${|\tilde i\rangle}$ of $g_{\lambda}$ are {\bf degenerate} with eigenvalue $a_{\lambda}$ for each term $\lambda$.

\ba
\hat g_{\lambda}|\tilde i\rangle=a_{\lambda}|\tilde i\rangle
\label{eq47}
\ea 
Then the total hamiltonian can be written

\ba
\hat H=\hat H_{S}+\hat H_{E}+\sum_{\lambda}a_{\lambda}\hat e_{\lambda}
\label{eq48}
\ea 
and the two last terms decouple evidently from $\hat H_{S}$ in the sense that now

\ba
e^{-it\hat H}=e^{-it\hat H_{S}(t}\otimes e^{-it(\hat H_{E}+\sum_{\lambda}a_{\lambda}\hat e_{\lambda})}
\label{eq49}
\ea  
i.e. the operators acting in $\cal H_{S}$ and $\cal H_{E}$ factorize. Since the system $S$ decouples from $E$ it is no longer affected by its environment and it stays coherent over time. This is a condition for the existence of so called decoherence free states in the subspace of $\cal H_{S}$ in which the decomposition given by eq.(46) works.\\

b) Model application\\

Consider the following system ~\cite{kr1}:

\begin{center}
\ba
\hat H_{S}=\omega \hat J_{z} 
\notag \\
\hat H_{E}=\beta b^{+}b
\notag \\
\hat H_{SE}=\eta(b^{+}+b) \hat J^{2}
\label{eq50}
\ea 
\end{center}  
where $b^{+},b$ are boson operators, $\omega$ is the rotation frequency of the system, $\beta$ the quantum of energy of the oscillator and $\eta$ the strength parameter in the coupling interaction between $S$ and $E$. Here $\hat C_{H_{SE}}=[\hat H_{S},\hat H_{SE}]=0$ trivially. Since $\hat J_{z}$ and $\hat J^{2}$ commute in the basis of states $[|j m\rangle]$ the projection of the total Hamiltonian $\hat H$ on $S$ space read
  
\ba
\langle jm|\hat H|jm \rangle=\omega m+\beta b^{+}b+\eta j(j+1)(b^{+}+b) 
\label{eq51}
\ea 
  
The expression of the density operator $\hat \rho_{S}(t)$ at time $t$ is then obtained by taking the trace over the environment states of the total Hamiltonian $\hat \rho(t)$ leading to 

\ba
\hat \rho_{S}(t)=Tr_{E}\hat \rho(t) 
\label{eq52}
\ea 
whose matrix elements read  
  
\ba
\rho^{j m_{1}, j  m_{2}}_{S}(t)=\rho^{j  m_{1}, j  m_{2}}_{0}(t)\Omega_{E}(j,j,t)
\label{eq53}
\ea 
with  
  
\ba
\rho^{j m_{1}, j m_{2}}_{0}(t)=\frac{e^{[-i\omega(m_{1}-m_{2})]t}}{(\hat j}
\label{eq54} 
\ea  
with $\hat j=2j+1$. The bosonic environment contribution can be put in the following form 
  
\ba
\Omega_{E}(j,j,t)=\sum_{n=0}^{n_{max}}\frac{1}{n!}\sum_{n',n^{"}}\frac{E_{n,n'}(j,t)
E^{*}_{n^{"},n}(j,t)}{[(n'!)(n''!)]^{1/2}}
\label{eq55} 
\ea  
These expressions are exact. The Zassenhaus development formulated in Appendix D has been used in order to work out the expressions ~\cite{zas}. The polynomials $E_{n,n'}(t)$ and $E^{*}_{n'',n}(t)$ are developed in Appendix E.\\

By simple inspection of the expressions in Appendix E it can be seen that the non-diagonal of $\rho^{j m_{1}, j m_{2}}_{S}(t)$ may cross zero when $t$ increases but oscillate and never reach and stay at zero whatever the length of the time interval which goes to infinity. Hence no decoherence will be observable in this case.

Compared with the preceding example the present interaction term contains a degeneracy of states of the system which explains the non-decoherent behaviour of the system ~\cite{lid1,agr}.
 
\section{Decoherence: summary} 
 
Coherence and decoherence are central concepts in quantum mechanics. They are the consequence of the wave nature of the microscopic world which induces the existence of interference effects in space and time and may lead to dephasing. 

In open stationary systems these effects can be absent under certain conditions if the Hamiltonian of the system $\hat H_{S}$ under consideration and the coupling Hamiltonian $\hat H_{SE}$ to its environment $E$ obey the commutation relation $[\hat H_{S},\hat H_{SE}]=0$. 

One may conjecture that the same condition is expected to work in the case of an open system which evolves in time. An example has been developed in section 6.1 in which an initial non-diagonal density matrix of the system goes over to a diagonal form with increasing time. 

How can this be understood? A possible explanation may come from the interpretation of the commutation relation between $\hat H_{S}$ and $\hat H_{SE}$. This relation constrains the system to stay in a fixed eigenstate of $\hat H_{S}$ if it is generated in this state at the origin of its coupling to an environment.

This is however not the case if several eigenstates of $\hat H_{S}$ are degenerate in 
$\hat H_{SE}$. In this case the coupling may not preclude interference effects between contributions of different basis states of $\hat H_{S}$. 

\section{Final Comments}

In the present work we aimed to analyze two essential properties of open quantum systems, Markovianity and coherence, starting from the general expression of the density operator of such systems. We showed how the examination of this expression leads to constraints which determine these  properties.  As expected their evolution in time are determined by the properties which govern the Hamiltonian structure of the systems, their environment and the coupling between the two. We worked them out and showed that both divisibility and coherence can, at least partially understood in terms of commutation relations involving the Hamiltonians of the system, the environment and their coupling.
We did not examine the case where the system obeys a Lindblad type equation ~\cite{gl,gk,pea,riv1}, a master equation  which obeys divisibility, linearity, hermiticity and complete positivity constraints. These constraints guarantee a decoherent behaviour of the  system. 

The present description concerned purely quantum systems and environments which are purely deterministic. The situation corresponding to the case where the environment can be treated in terms of quantum statistics has been introduced recently by Allahverdyan, Balian and Nieuwenhuizen~\cite{al1,al2} in the framework of a measure theory in which the environment is a macroscopic system. The approach based on a statistical treatment of a macroscopic environment considers a different situation which wants to provide an answer to the measure problem. However this is a vast and not yet completely understood question
~\cite{zeh,whe,zur1,zur2,per,kie,sch}.

\section{Appendix A: imposing the divisibility constraint}

Using the explicit expression of the super matrix $C$ given by  Eq.(17) the divisibility constraint in Eq.(20) for fixed states $(i_{1}, i_{2})$, $(k_{1}, k_{2})$ imposes the following relation 
 
\ba
\sum_{\alpha_{1},\alpha_{2},\gamma}d_{\alpha_{1},\alpha_{2}}U_{(i_{1}k_{1}),(\alpha_{1}\gamma)}(t-t_{0})
U^{*}_{(i_{2}k_{2}),(\alpha_{2}\gamma)}(t-t_{0})=
\sum_{j_{1},j_{2}}\sum_{\alpha_{1},\alpha_{2},\beta_{1},\beta_{2}}d_{\alpha_{1},\alpha_{2}} 
d_{\beta_{1},\beta_{2}}
\notag \\
\sum_{\gamma,\delta}U_{(j_{1}k_{1}),(\beta_{1}\delta)}(t-t_{s})U_{(i_{1}j_{1}),(\alpha_{1}\gamma)}(t_{s}-t_{0})
U^{*}_{(j_{2}k_{2}),(\beta_{2}\delta)}(t-t_{s})U^{*}_{(i_{2}j_{2}),(\alpha_{2}\gamma)}(t_{s}-t_{0}) 
\label{eq56}
\ea

In order to find a solution to this equality and without loss of generality we consider the case where the density matrix in $E$ space is diagonal. Then the equality reads 
 
\ba
\sum_{\alpha, \gamma}d_{\alpha,\alpha}U_{(i_{1}k_{1}),(\alpha \gamma)}(t-t_{0})
U^{*}_{(i_{2}k_{2}),(\alpha \gamma)}(t-t_{0})=
\sum_{j_{1},j_{2}}\sum_{\alpha,\beta}d_{\alpha,\alpha} d_{\beta,\beta}
\notag \\
\sum_{\gamma,\delta}U_{(j_{1}k_{1}),(\beta \delta)}(t-t_{s})U_{(i_{1}j_{1}),(\alpha \gamma)}(t_{s}-t_{0})
U^{*}_{(j_{2}k_{2}),(\beta \delta)}(t-t_{s})U^{*}_{(i_{2}j_{2}),(\alpha \gamma)}(t_{s}-t_{0}) 
\label{eq57}
\ea

A sufficient condition to realize the equality is obtained if $d_{\beta,\beta}=d_{\alpha,\alpha}$ and consequently if the weights $d$ on both sides are to be the same one ends up with $d_{\alpha,\alpha}=1$. This last condition imposes a unique state in $E$ space, say $|\eta \rangle$. In this case $d_{\eta,\eta}=1$ and Eq.(56) reduces to  
 
\ba
U_{(i_{1}k_{1}),(\eta \eta)}(t-t_{0})U^{*}_{(i_{2}k_{2}),(\eta \eta)}(t-t_{0})=
\sum_{j_{1}} U_{(i_{1}j_{1}),(\eta \eta)}(t_{s}-t_{0})U_{(j_{1}k_{1}),(\eta \eta)}(t-t_{s})
\notag\\
\sum_{j_{2}} U^{*}_{(j_{2}k_{2}),(\eta \eta)}(t-t_{s})U^{*}_{(i_{2}j_{2}),(\eta \eta)}(t_{s}-t_{0}) 
\label{eq58}
\ea
which proves the equality.

\section{Appendix B: a special case of divisibility}  

Starting from the expression of the density operator given by Eqs.(15-18) we consider the case where $|c_{i}|=1/n$  for all $i$ where $n$ is the number of states in $S$ space and $d_{\alpha_{1},\alpha_{2}}=1/N \delta_{\alpha_{1},\alpha_{2}}$.\\

In this case the relation which imposes the divisibility constraint reads 
 
\ba
\frac{1}{Nn}\sum_{i\alpha,\gamma}U_{(ik_{1}),(\alpha\gamma)}(t,t_{0})U^{*}_{(ik_{2}),(\alpha\gamma)}(t,t_{0})=
\notag\\ 
\frac{1}{N^{2}n}\sum_{j_{1}j_{2}\beta,\delta}U_{(j_{1}k_{1}),(\beta\delta)}(t,t_{s})U^{*}_{(j_{2}k_{2}),(\beta\delta)}(t,t_{s})
\notag\\
\sum_{i,\alpha,\gamma}U_{(ij_{1}),(\alpha\gamma)}(t_{s},t_{0})U^{*}_{(ij_{2}),(\alpha\gamma)}(t_{s},t_{0})
\label{eq59} 
\ea

The expression in the last line leads to  

\ba
\sum_{i,\alpha,\gamma}U_{(ij_{1}),(\alpha\gamma)}(t_{s},t_{0})U^{*}_{(ij_{2}),(\alpha\gamma)}(t_{s},t_{0})= N\delta_{j_{1},j_{2}} 
\label{eq60}
\ea 
 
and finally the r.h.s. reduces to

\ba 
\sum_{j_{1}j_{2}\beta,\delta}U_{(j_{1}k_{1}),(\beta\delta)}(t,t_{s})U^{*}_{(j_{2}k_{2}),(\beta\delta)}(t,t_{s})=1/N\delta_{k_{1},k_{2}} 
\label{eq61}
\ea 
 
It is easy to see that working out the l.h.s. of Eq.(59) leads to the same result.  

\section{Appendix C: general case of divisibility} 

For a unique fixed state $\gamma$
the expressions of $\frac{d}{dt}\rho_{S1 \gamma}^{j_{1}j_{2}}(t)$ and $\frac{d}{dt}\rho_{S2 \gamma}^{j_{1}j_{2}}(t)$ given in 
Eqs.(21-22) can be written as
 
\ba
\frac{d}{dt}\rho_{S1 \gamma}^{j_{1}j_{2}}(t)=(-i)\sum_{k_{1}k_{2}}
A^{j_{1}k_{1}}_{\gamma}\rho^{k_{1}k_{2}}_{S1 \gamma}(t)I^{k_{2}j_{2}} 
\label{eq62}
\ea
where $\hat I$ is the identity operator in S space and 
 
\ba
A^{j_{1}k_{1}}_{\gamma}=\langle j_{1}\gamma|\hat H_{S}|k_{1}\gamma\rangle + \langle j_{1}\gamma|\hat H_{E}+\hat H_{SE}|k_{1}\gamma\rangle
\label{eq63}
\ea 
and similar expressions for $d/dt \rho_{S2 \gamma}^{j_{1}j_{2}}(t)$. The matrix elements of $\hat H_{SE}$ in the second term on the r.h.s. of the expression of $A$ are generally non diagonal in $E$. They are diagonal if $\hat H_{E}$ and $\hat H_{SE}$ commute in the basis of states in which $\hat H_{E}$ is diagonal.

In symmetrized form the r.h.s. of the master equation obtained from Eq.(24) reads 

\ba
(-i)\hat H \hat\rho_{S} +(i)\hat\rho_{S}\hat H=(\hat I-i\hat H)\hat\rho_{S}(\hat I+i\hat H)
-\hat\rho_{S}-\hat H \hat\rho_{S}\hat H
\label{eq64}
\ea 

\section{Appendix D: the Zassenhaus development} 
 
If 
$X=-i(t-t_{0})(\hat H_{S}+\hat H_{E})$ and $Y=-i(t-t_{0})\hat H_{SE}$

\ba
e^{X+Y}=e^{X}\otimes e^{Y}\otimes e^{-c_{2}(X,Y)/2!}\otimes e^{-c_{3}(X,Y)/3!}\otimes e^{-c_{4}(X,Y)/4!}...
\label{eq65}
\ea

where

\begin{center}
$c_{2}(X,Y)=[X,Y]$\\ 
$c_{3}(X,Y)=2[[X,Y],Y]+[[X,Y],X]$\\ 
$c_{4}(X,Y)=c_{3}(X,Y)+3[[[X,Y],Y],Y]+[[[X,Y],X],Y]+[[X,Y],[X,Y]$\\
\end{center} 
 
The series has an infinite number of term which can be generated iteratively in a straightforward way ~\cite{ca}. If $[X,Y]=0$ the truncation at the  third term  leads to the factorisation of the $X$ and the $Y$ contribution. If $[X,Y]=c$ where $c$ is a c-number the expression corresponds to the well-known Baker-Campbell-Hausdorff formula.
  
\section{Appendix E: Expression of the bosonic sector} 

The expressions of the bosonic contributions to the density matrix $\rho^{j_{1} m_{1}, j_{2} m_{2}}_{s}(t)$ are given by 
 
\ba
E_{n,n'}(j_{1},t)=e^{-i\beta t}\sum_{n\geq n_{2},n_{3}\geq n_{2}}\sum_{n_{3}\geq n_{4},n'\geq n_{4}}(-i)^{n+n_{3}}
(-1)^{n'+n_{2}-n_{4}}
\notag\\
\frac{n!n'!(n_{3}!)^{2}[\alpha(t)^{n+n_{3}-2n_{2}}][\zeta(t)^{n_{3}+n'-2n_{4}}]}{(n-n_{2})!(n_{3}-n_{4})!
(n_{3}-n_{2})!(n'-n_{4})!}e^{\Psi_{1}(t)}
\label{eq66}      
\ea
and

\ba
E^{*}_{n^{"},n}(t;j_{2})=e^{i\beta t}\sum_{n^{"}\geq n_{2},n_{3}\geq n_{2}}\sum_{n_{3}\geq n_{4},n\geq n_{4}}i^{n^{"}+n_{3}}
(-1)^{n+n_{2}-n_{4}}
\notag\\
\frac{n^{"}!n!(n_{3}!)^{2}[\alpha(t)^{n^{"}+n_{3}-2n_{2}}][\zeta(t)^{n+n_{3}-2n_{4}}]}{(n^{"}-n_{2})!(n_{3}-n_{2})!(n_{3}-n_{4})!(n-n_{4})!}e^{\Psi_{2}(t)}
\label{eq67}      
\ea

The different quantities which enter $E_{n,n'}(t)$ are 

\ba
\alpha(t)=\frac{\gamma(j_{1})\sin\beta t}{\beta}
\label{eq68}      
\ea

\ba
\zeta(t)=\frac{\beta[1-\cos\gamma(j_{1})t]}{\gamma(j_{1})}
\label{eq69}      
\ea

\ba
\gamma(j_{1})=\eta j_{1}(j_{1}+1)
\label{eq70}      
\ea

\ba
\Psi_{1}(t)=-\frac{1}{2}[\frac{\gamma^{2}(j_{1})\sin^{2}(\beta t)}{\beta^{2}}+\frac{\beta^{2}(1-\cos\gamma(j_{1})t)^{2}}{\gamma^{2}(j_{1})}]                           
\label{eq71}      
\ea

and similar expressions with index 2 for $E^{*}_{n'',n}(t)$. In practice $j_{1}=j_{2}=j$ in section 6.2.


\begin{thebibliography}{99}

\bibitem{riv1} \'Angel Rivas, Susana F. Huelga and Martin B. Plenio, Rep.Prog. Phys. 77 (2014) 094001

\bibitem{pol} Felix A. Pollock, C\'esar Rodr\'iguez-Rozario, Thomas Frauenheim, Mauro Paternostro 
and Kavan Modi, arXiv:1512.00589 [quant-ph]

\bibitem{ad} S.A. Adelman, J. Chem. Phys. 64 (1976) 124

\bibitem{fle1} C.H. Fleming, Albert Roura, B.L. Hu, Ann. Phys. 326 (2011) 1207

\bibitem{sp} Herbert Spohn, Rev. Mod. Phys. 53 (1980) 569 

\bibitem{gk} V. Gorini, A. Kossakowski, E.C.G. Sudarshan, J. Math. Phys. 17 (1976) 821

\bibitem{gl} G. Lindblad, Comm. Mat. Phys. 48 (1976) 119

\bibitem{lzb} Daniel A. Lidar, Zsolt Bihari, K. Brigitta Whalley, Chem. Phys. 268 (2001) 35

\bibitem{sr} T. Sami and J. Richert, Z. Phys. A- Atoms and Nuclei 317 (1984) 101
 
\bibitem{hal} Michael J. W. Hall, James D. Cresser, Li Li, Erika Andersson, Phys. Rev. A 89 (2014) 042120 and refs. therein
 
\bibitem{hb} Heinz-Peter Breuer, Phys. Rev. A 75 (2007) 022103
 
\bibitem{vb} Bassano Vacchini, Heinz-Peter Breuer, Phys. Rev. A 81 (2010) 042103  
 
\bibitem{ar} \'Angel Rivas, Susana F. Huelga, Martin B. Plenio, Phys. Rev. Lett. 105 (2010) 050403
 
\bibitem{fb} Luca Ferialdi, Angelo Bassi,  Phys. Rev. Lett. 108 (2012) 170404  

\bibitem{sl} Salvatore Lorenzo, Francesco Plastina, Mauro Paternostro, Phys. Rev. A 84 (2011) 032124 

\bibitem{ss} Shunlong Luo, Shuangshuang Fu, Hongting Song, Phys. Rev. A 86 (2012) 044101 

\bibitem{mw} M.M. Wolf, J. Eisert, T.S. Cubitt and J.I. Cirac, Phys. Rev. Lett. 101 (2008) 150402 

\bibitem{bv} Bassano Vacchini, Phys. Rev. A 87 (2013) 030101(R)

\bibitem{blp} Heinz-Peter Breuer, Elsi-Mari Laine, Jyrki Piilo, Phys. Rev. Lett. 103 (2009) 210401 

\bibitem{afp} Tony J. G. Apollaro, Carlo Di Franco, Francesco Plastina, Mauro Paternostro, Phys. Rev. A 83 (2011) 032103
 
\bibitem{pcz} J. F. Poyatos and J. I. Cirac, P.Zoller, Phys. Rev. Lett. 78 (1997) 39 
 
\bibitem{hpb} Heinz-Peter Breuer, J. Phys. B: At. Mol. Opt. Phys. 45 (2012) 154001 

\bibitem{ck} Dariusz Chru\'si\'nski and Andrzej Kossakowski, Eur. Phys. J. D (2014) 68: 7 
 
\bibitem{mark} R. Howard, Dynamic Probabilistic Systems, vol. 1, Appendix B: Markov Chains, 1971,
John Wiley and Sons eds.

\bibitem{oli} Oliver C. Ibe, Markov Processes for Stochastic modeling, 2nd edition, Elsevier ed.

\bibitem{har} S. Haroche, cours du Coll\`ege de France, chaire de Physique, Ann\'ee 2003 - 2004 - 4\`eme cours 

\bibitem{vlb} Vladimir Bu\v{z}ek, Phys. Rev. A 58 (1998) 1723

\bibitem{has} S. Haseli, G. Karpat, S. Salimi, A. S. Khorashad, F. F. Franchini, B. Cakmak, G. H. Aguilar, S. P. Walborn, and P.H. Souto Ribeiro, Phys. Rev. A90 (2014) 052118

\bibitem{sti} W. F. Stinespring, Proc. Amer. Math. Soc. 6 (1955) 211

\bibitem{bau} T. Baumgratz, M. Cramer and M.B. Plenio, Phys. Rev. Lett. 113 (2014) 140401

\bibitem{nak} S. Nakajima, "On Quantum Theory of Transport Phenomena",  Prog. Theor. Phys. (1958) 20 (6): 948-959

\bibitem{zwa} R. Zwanzig, "Ensemble Method in the Theory of Irreversibility", J. Chem. Phys. 33 (1960) 1338

\bibitem{ck1} Dariusz Chru\'si\'nski and Andrzej Kossakowski, Phys. Rev. Lett. 111 (2013) 050402

\bibitem{rei} John H. Reina, Luis Quiroga and Neil F. Johnson, Phys. Rev. A 65 (2002) 032326

\bibitem{add} Carole Addis, Francesco Ciccarello, Michele Cascio, G. Massimo Palma and Sabrina Maniscalco, New J. Phys. 17 (2015) 123004

\bibitem{lid1} D.A. Lidar, I.L. Chuang and K.B. Whaley,  Phys. Rev. Lett. 81 (1998) 2594

\bibitem{agr} Julian Agredo, Franco Fagnola and Rolando Rebolledo, J. Math. Phys. 55 52014) 112201 
 
\bibitem{kr1} T. Khalil and J. Richert, arXiv:1503.08920 [quant-ph]
 
\bibitem{zas} H. Zassenhaus, Abh. Math. Sem. Univ. Hamburg 13 (1940) 1 - 100  
 
\bibitem{pea} Philip Pearle, Europ. J. Phys. 33 (2012) 805

\bibitem{al1} A. E. Allahverdyan, R. Balian and T. H. Nieuwenhuizen, Physics Reports {\bf525}, (2013) 1

\bibitem{al2} A. E. Allahverdyan, R. Balian and T. H. Nieuwenhuizen, arXiv:1303.7257v1 [quant-ph]
  
\bibitem{zeh} Dieter Zeh, Found. of Phys. (1970) 69

\bibitem{whe} J.A. Wheeler and W.H. Zurek, "Quantum Theory of Measurement", Princeton University Press, 1983

\bibitem{zur1} W.H. Zurek, I.N. Perez-Mercader, J. Halliwell and W. Zurek editors, "Physical Origins of Time Asymmetry", Cambridge University Press, 1994 

\bibitem{zur2} W.H. Zurek, Rev. Mod. Phys. 75 (2003) 715

\bibitem{per} A. Peres, "Quantum Theory: concepts and methods", Kluwer Academic Publishers, 1995

\bibitem{kie} C. Kiefer, J. Kupsch, I.O. Stamatescu, D. Giulini, E. Joss and W.H. Zurek, "Decoherence and the appearance of a classical world",  Springer Berlin 1996

\bibitem{sch} M. Schlosshauer, "Decoherence and the Quantum to Classical Transition", Springer Heidelberg Berlin, 2007
 
\bibitem{ca} Fernando Casas, Ander Murua, Mladen Nadinic, Computer Physics Communications 183 (2012) 2386







































 
 
 

 
 
  

\end{thebibliography}
\end{document}